\documentclass[aps,prb,a4paper]{revtex4}
\usepackage[dvips]{graphicx}

\begin{document}

\title{Predicting protein functions with message passing algorithms}

\author{M. Leone$^1$, A. Pagnani$^1$} 
 
\affiliation{$^1$ Institute for Scientific Interchange (ISI), Viale
  Settimio Severo 65, I-10133 , Turin, Italy}

%\date{\today}

%\maketitle

\begin{abstract}

{\bf Motivation: } In the last few years a growing interest in biology has been
shifting towards the problem of optimal information extraction from
the huge amount of data generated via large scale and high-throughput
techniques. One of the most relevant issues has recently become that
of correctly and reliably predicting the functions of observed but
still functionally undetermined proteins starting from information
coming from the network of co-observed proteins of known functions
~\cite{CIT-ALEXEI1}.

\noindent 
{\bf Method: } The method proposed in this article is based on a
message passing algorithm known as Belief Propagation
~\cite{YEDIDIA}, which takes as input the network of proteins physical
interactions and a catalog of known proteins functions, and returns
the probabilities for each unclassified protein of having one
chosen function. The implementation of the algorithm allows for fast
on-line analysis, and can be easily generalized to more complex graph
topologies taking into account hyper-graphs, {\em i.e.} complexes of
more than two interacting proteins. 

\noindent
{\bf Results: } The benchmark of our method is the
{\it Saccaromices Cerevisi\ae } protein-protein interaction network
(PPI)~\cite{UETZ,DIP} and the validity of our approach is successfully
tested against other available
techniques~\cite{OtherFUNCTION-Methods,BIOINFO,ALEXEI}.

\noindent
{\bf Contact: } leone@isiosf.isi.it, andrea.pagnani@roma1.infn.it


\end{abstract}

\maketitle

\section{Introduction}

The most classical protein function prediction
methods are those inferring similarity in function from sequence
homologies between proteins listed in databases using programs such as
FASTA~\cite{FASTA} and BLAST~\cite{BLAST}; via comparison with known
proteins interactions in similar genomes (the so called {\it Rosetta
Stone Method}~\cite{Rosetta}); or by phylogenetic analysis~\cite{Phylo}. 
More recently, a new class of methods has been proposed
that relies on the available data on the global structure of the PPI
networks for a growing number of organisms of completely sequenced
genome~\cite{UETZ,ITO,DROSOPHILA}.  The most complete available
on-line data are structured in a graph-like format, with graph sites
indexed with protein names and links representing a physical
experimentally tested interaction among two proteins. More limited
databases on larger protein complexes are also 
available~\cite{COMPLEXES1,COMPLEXES2}. 
From the side of functional
classification, databases are now available 
({\bf MIPS}\footnote{The MIPS Comprehensive Yeast Genome Database (CYGD),
  http://mips.gsf.de/proj/yeast/CYGD/db/.} and {\bf Gene Ontology}
among others\footnote{The Gene Ontology Consortium, http://www.geneontology.org/}), 
that provide a classification of a continuously growing number of proteins,
listing them in different functional categories classes with a
hierarchical-like organization.  Among the presently available methods
that try to exploit the global PPI network structure to infer yet
unknown functions for unclassified proteins whose interactions with the
rest of the graph are at least partially known, there are the so
called {\it Neighboring Counting Method}~\cite{UETZ2}, the {\it
$\chi^2$ Method}~\cite{CHI}, the Bayesian 
approaches~\cite{OtherFUNCTION-Methods,BIOINFO}, the 
{\it Redundancy Method}~\cite{LIANG} and a more recent
Monte Carlo Simulated Annealing (SA) approach~\cite{ALEXEI}.

\section{Methods} 

\begin{figure}
\includegraphics[width=3.5in]{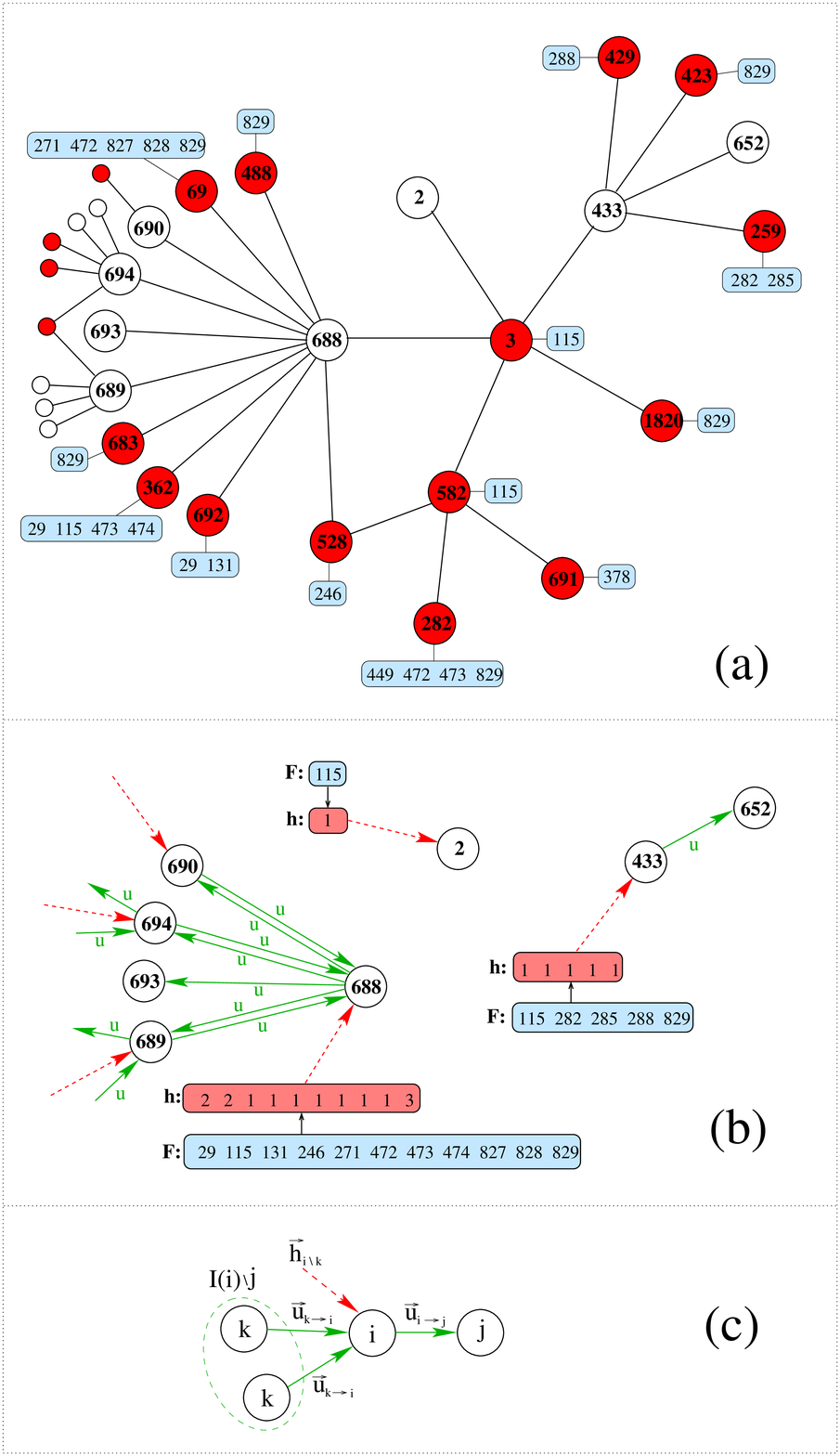}
\caption{ From ${\cal G}$ to the BP
equations: fig.{\bf (a)} shows a small fraction of {\bf U} network 
${\cal G}$. Circles represent proteins with
their numerical ID used by the algorithm. Classified proteins are
filled, while unclassified ones are left white. Each classified
protein has a series of functions whose numerical values $\in$ ${\cal
F}$ are written in boxes. In {\bf (b)} only the
corresponding part of the ${\cal U}$ subgraph has been drawn.
Dotted arrows represent external fields acting on the unclassified
proteins and are vectors whose non zero components are defined in the
Lower boxes. For each protein in ${\cal U}$, they count the number
of classified neighbors having a given function. Upper boxes are
sets of all functions of all classified proteins neighboring a given
unclassified one. Thick arrows represent ``messages'' 
among unclassified proteins according to
eq.~(\ref{eq:M}). Notice how ${\cal U}$ is significantly less
connected than ${\cal G}$ and often divides in smaller connected
components. {\bf (c)} Is a more detailed representation of the message passing 
between proteins $i$ and $j$, in direction $i \to j$. }
\label{fig:graph}
\end{figure}

Let us name ${\cal G}$ a PPI graph, with set of vertexes $V =\{1,\cdots ,N\}$
representing the observed proteins, each protein name being assigned a
numerical value form $1$ to $N$. Let us also define a mapping between
the set of all observed functions and the numbered set ${\cal F} =
\{1,\cdots,F\}$. Each protein $i$ belonging to $V$ can then be
characterized via a discrete variable $X_i$ that can take values $f
\in {\cal F}$.  One would like to compute the probability $P_i(f) =
Pr(X_i = f)$ for each protein to have a given function $f$ given the
functions assigned to the proteins in the rest of the graph.  The
method is based on the definition of a score function $E$ on
the PPI graph (see eq.~(\ref{score})), that counts
the number of all common predicted functions among neighboring
proteins of the graph over all interactions.  In addition to this, a
certain fraction of the proteins is already classified, which means
that there exists a subset $A \subset V$ of vertexes with at least one
function belonging to ${\cal F}$ attached to it (see fig.(\ref{fig:graph})
{\bf(a)} for an example of a graph portion). The effect of the
already classified proteins with a given function in the neighborhood
of protein $i$ on the PPI network is taken into account as an external
field acting on $i$ and proportional to the number of the neighbors
belonging to $A$ with that given function. From this score function a
variational potential (called Gibbs potential) can be defined that
measures the distance between the true unknown function probabilities
and a trial estimation of them. The values of the best estimated
probabilities are found extremizing the Gibbs 
potential~\cite{YEDIDIA,JAAKKOLA}. The Gibbs potential extremizing equations
used in this work are commonly known under the name of {\it Belief
Propagation} (BP) equations and can be easily found via a procedure
called {\it Cavity Method}~\cite{BETHE-REVISITED}. We have solved the
BP equations both for the probabilities of completely unclassified
proteins belonging to $V \backslash A$ and for the more complete model
where we let a protein belonging to $A$ the possibility of having
other yet unknown functions. The modifications to be applied to the
method are technical minor so that they will not be described here.
Given a
choice of initial conditions on probability functions $\{ P_i(X_i)
\}_{i=1,...,N}$ and a choice of the score function $E$, the algorithm
calculates the stationary probabilities whose values extremize the
resulting Gibbs potential. The potential in general depends on one
free real parameter $\beta$ that plays the role of an inverse
temperature and weights the possibility of allowing functional
assignments that do not exactly maximize the score, but could still be
possible due to their large degeneracy: at low enough values of
$\beta$ (high temperature) almost any function assignation to proteins
in $V \backslash A$ gives and equivalent value of the potential. In
this region the system is said to be in a ``paramagnetic
phase''. Every functional assignment is therefore accepted and the
algorithm is not predictive. After a certain critical value $\beta_c$
the shape of the Gibbs potential changes: only some values of the
probability functions extremize it. Augmenting $\beta$, the algorithm
tends to weight more and more those functional assignments that
exactly maximize the score. Strictly at zero temperature ($ \beta \to
\infty$) only the score maximizing functional assignments survive with
non zero probability.
Given sets $V$, $A$ and $V \backslash A$, the PPI graph ${\cal G}$, 
the graph of unclassified proteins
${\cal U} \subset {\cal G}$ and the set of observed function ${\cal
F}$, a score function can be defined following Vazquez et al.~\cite{ALEXEI} as
\begin{equation}
E[\{X_i\}_{i=1}^N] = -\sum_{ij} J_{ij} \delta ( X_i;X_j ) 
- \sum_i h_i (X_i)
\label{score}
\end{equation}
where $J_{ij}$ is the adjacency matrix of ${\cal U}$ ($J_{ij} = 1$ if
$i$ and $j$ $\in V \backslash A$ and they interact with each
other). $\delta( \sigma; \tau )$ is the Kronecker delta function
measured between functions $\sigma$ and $\tau$ assigned to the
neighboring proteins and $ h_i (\tau) $ is an external field that
counts the number of classified neighbors of protein $i$ in the
original graph ${\cal G}$ that have at least function $\tau$.  The
Gibbs potential can than be calculated as a variational way to compute
the quantity
\begin{equation}
F = -\frac{1}{\beta N}\log \left( \sum_{\{X_i\}_{i=1,...,N}} e^{-\beta
  E[\{X_i\}_{i=1,...,N}] } \right) 
\label{Potential}
\end{equation}
called free-energy of the system, a fundamental quantity than in
statistical physics counts the logarithm of the sum of all the weights
of the probabilities each configuration of the variables in the
systems appear with.  Configurations with a largest statistical weight
can then be calculated as those maximizing this potential function.
Using the message passing approach~\cite{YEDIDIA,BETHE-REVISITED} under the
assumption that correlations are low
enough in the graph so that one can write $P_{ij}(X_i,X_j) \propto
P_i(X_i)P_j(X_j)$ if proteins $i$ and $j$ are chosen at random,
one can calculate each $P_j(X_j)$ as product of conditional
probabilities contributions $M_{i \to j}(X_j)$ incoming to $j$ from
all neighbors of protein $j$, conditional to the fact that $j$ has
function $X_j$:
\begin{equation}
P_j(X_j) \propto \prod_{i \in I(j)} M_{i \to j}(X_j) = e^{\beta \sum_{i \in
    I(j)} u_{i \to j}(X_j)}
\label{eq:prob}
\end{equation}
where $I(j) \subset V \backslash A$ denotes the set of unclassified
neighbors of $j$ and $u_{i \to j}(\sigma)$ is a ``message'' that
represents the field in direction $\sigma \in {\cal F}$ acting on
protein $j$ due to the presence of protein $i$ when protein $j$ has
function $\sigma$. Equations for the message functions can be solved
iteratively as fixed points of the system of equations
\begin{equation}
M_{i \to j}(\sigma) = \sum_{\tau = 1}^F \left( \prod_{l \in I(i)\backslash
  j} M_{l \to i}(\tau) \right)
e^{\beta J_{ij}\delta (\sigma; \tau) + \beta h_i (\tau)}
\label{eq:M}
\end{equation}
one for each link of ${\cal U}$, for both directions in the graph.  
Self consistent BP equations can be rewritten in terms of messages $u$'s. The ones 
explicitly used in our algorithm are shown in the following:
\begin{equation}
u_{i \to j}(\sigma) = \frac{1}{\beta} \log \left(
\frac{A(\beta ; {\vec T}^1_{i \to j} (\sigma) )}{A(\beta ; {\vec T}^2_{i \to j})}
\right)
\label{eq:u}
\end{equation}
where
\begin{displaymath}
\left\{
\begin{array}{lll}
A(\beta; {\vec T}_{i \to j}) &=& \sum_{\tau = 1}^F e^{\beta T^{\tau}_{i
    \to j}} \\
\vec{T}^{1,\tau}_{i \to j} (\sigma) &=& h_i(\tau) + \sum_{l \in I(i) \backslash j} u_{l
 \to i}(\tau) \\
&if& \sigma \neq \tau \;\; \\
&or&  \;\; h_i(\tau) + \sum_{l \in I(i) \backslash j} u_{l
 \to i}(\tau) + 1 \\
&if& \;\; \sigma = \tau \\
\vec{T}^{2,\tau}_{i \to j} &=& h_i(\tau) + \sum_{l \in I(i) \backslash j} u_{l
 \to i}(\tau) \\
\end{array}
\right.
\end{displaymath}
The BP algorithm has been written in terms of
that equation and solved at any $\beta$ with a population dynamic
technique~\cite{BETHE-REVISITED}.  In general, all
previously described quantities depend on the inverse temperature
$\beta$.  Eq.~(\ref{eq:prob}) turns out to be a good approximation of
the solution of the problem of finding the probabilities for
configurations maximizing (\ref{Potential}).
A pictorial view of the iteration procedure is shown in
fig~(\ref{fig:graph})(c).

\section{Data and graph analysis} 

As benchmarks for the method, we have used two yeast {\it Saccaromices
Cerevisi\ae } PPI graphs~\cite{UETZ,DIP}, referred in the following as 
{\bf U} and {\bf D} respectively. The functional categories set ${\cal F}$ was
extracted form the MIPS database. 
The {\bf U} network contains $N = 1826$ proteins out of which
$1370$ belong to $A$, while the remaining $456$ are unclassified or
have an unclear MIPS classification; and $M = 2238$ pairwise
interactions.  The {\bf D} network contains $N = 4713$ proteins out
of which $3303$ belong to $A$ and $1410$ are unclassified; and $M =
14846$ interactions. Different choices of
functional categories sets ${\cal F}$ are possible in the MIPS database,
depending on the level of the coarse-grained specification of the
hierarchical classification scheme. We used the latest publicly
available finest classification scheme retrieving $F = 165$ functional
categories present {\bf U} and $F = 176$ in {\bf D}, but
experiments where run also on the most coarse-grained classification
scheme. Results are available upon request. 

The PPI graph consists of a giant component of 1299 sites (990
classified), and the rest of the sites are grouped into 184 smaller
isolated components of at most 13 sites. We have also analyzed the
structure of the $V\backslash A$ graph which turns of
456 sites, grouped in 309 clusters of size at most 27. 
Each cluster in $V\backslash A$ can
be considered as an isolated {\em functional island} of the graph
surrounded by external fields as displayed in the last picture of 
the main body of the paper. 
More details on the cluster composition of both  ${\cal G}$ and
${\cal U}$ for the {\bf U} PPI network are shown in Table \ref{tab:graph}. 
One may wonder if these clusters are more then a
topological feature of our model, but reflect also a more interesting
{\em functional segregation}. In other words one is interested to
understand in quantitative terms how different clusters in
$V\backslash A$ label different functional areas in our graphs. To
this aim we measured inter-cluster and intra-cluster functional
overlaps as in eqs.~(\ref{overlap_intra}) and (\ref{overlap_inter}) 
Both observables take value in the interval $(0,1)$ and
give a measure of the functional similarity of clusters (higher values
indicate higher similarity). The emerging scenario shows clear signs
of segregation since the intra-cluster overlap distributions has
support in the interval $(0,0.1)$ while the inter-cluster distribution
has support in the whole interval $(0,1)$. This test can be
interpreted as a coherence test on the graph itself, and also on the
working hypothesis of our method, since segregation is tacitly
assumed in the functional form of score function where
only first neighbors interations on the graph are taken into account.
\begin{table}
\begin{center}
\begin{tabular}{cccc}
\hline
\multicolumn{2}{c}{${\cal G}$ } & \multicolumn{2}{c}{${\cal U}$ }\\
\hline
cs & noc& cs & noc\\
\hline 
2 & 114 & 1 & 248\\
3 & 30 & 2 & 40\\
4 & 23 & 3 & 8\\  
5 & 6 &	4 & 7\\  
6 & 4 & 5 &1 \\  
7 & 1 & 6 & 1\\  
8 & 4 & 7 & 1\\  
11& 1 & 14 & 1 \\
13 & 1 & 17 & 1 \\
1299 & 1 & 27 & 1\\ 
\hline
\end{tabular}
\caption{Cluster size (cs) and number of clusters (noc) for both the
  original graph ${\cal G}$ (the two leftmost columns) and the graph of
  the unknow proteins ${\cal U}$ (the two rightmost columns). Note the
  giant componenent of 1299 sites for the ${\cal G}$ graph. }
\label{tab:graph}
\end{center}
\end{table}
Let us define the notion of intra and inter cluster functional overlap as 
\begin{eqnarray}
\label{overlap_intra}
O_i & = & \frac 1 {N_i^2} \sum_{l,k \in {\cal C}_i} \
\frac {\phi( s_l, s_l ) } { \Phi({\cal C}_i) }\\
\label{overlap_inter}
o_{ij} &=& \frac 1 {N_i N_j } \sum_{l\in {\cal C}_i} \sum_{l\in {\cal C}_j}
\frac {\phi(s_l,s_k)}{\Phi({\cal C}_i \bigcup  {\cal C}_j ) }\,\,\,\,\,\,\, i\neq j
\end{eqnarray}
where index $i$ labels the different clusters ${\cal C}_i $ and run
between 1 and the total number of clusters $C$, $N_i$ is the number of
site in cluster ${\cal C}_i$, $\phi(s_l,s_k)$ counts the number of
function that site $l$ and $j$ have in common, and $\Phi({\cal C}_i)$
is the number of different functions acting onto cluster ${\cal C}_i$,
while $\Phi({\cal C}_i \bigcup {\cal C}_j )$ is the number of
different functions acting onto the union set ${\cal C}_i \bigcup
{\cal C}_j$. It is interesting to note that according to
Eqs.~\ref{overlap_intra},~\ref{overlap_inter} both $o_i$ and $O_{ij}$
have take real values in the interval $(0,1)$.  We can consider probability
densities of the two variables $O_i$ and $o_{ij}$ as 
\begin{eqnarray}
P^{(\mathrm{intra})} (O) &=& \frac 1N \sum_{i=1}^C \delta(O,O_i)\\
P^{(\mathrm{inter})} (o) &=& \frac 1{N(N-1)} 
\sum_{1\leq i<j\leq C} \delta(o, o_{ij})
\end{eqnarray}
where $\delta (a, b) $ is the Kronecker delta function equal to 1 when
$a=b$ and zero otherwise. It is interesting a this point to compare
the average intracluster overlap $\langle O \rangle = 0.440673$ with
the average entercluster overlap $\langle o \rangle = 0.0294147$ which
is a factor 15 smaller and that can be taken as a quantitative
measure of the functional segregation on the PPI graph.

\begin{figure}
\begin{center}
\includegraphics[angle=0,width=0.95\columnwidth]{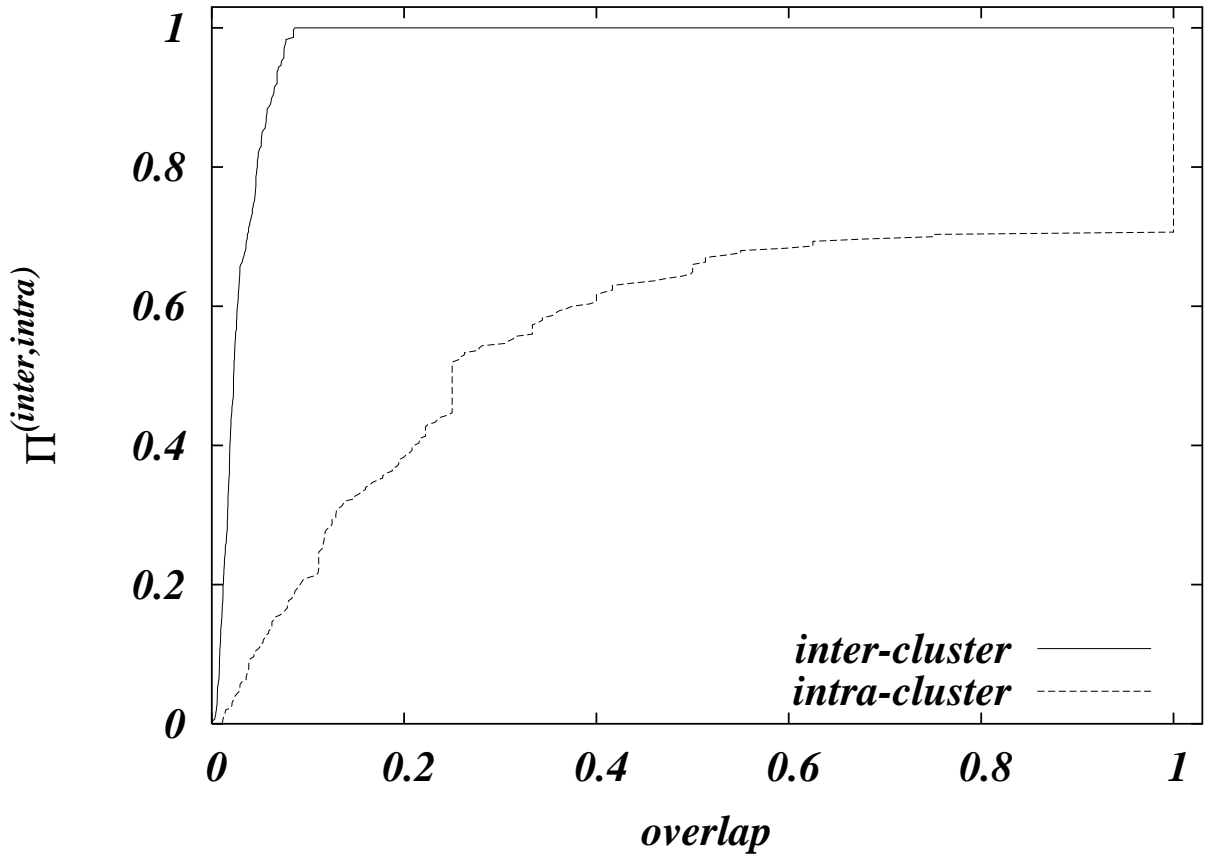}
\caption{Cumulative probability distribution of intra/inter-cluster
funtional overlap as defined in
Eqs.~\ref{overlap_intra},~\ref{overlap_inter} for the {\bf U} ${\cal U}$ graph . Since the intra-cluster
overlap turns out to be alwais lower than 0.085, the intra-cluster
cumulative probability distribution (solid line) saturates to 1 above
this value. The inter-cluster overlap instead shows clear sign of
segregation. Note that the sudden jump at 1 for the dashed line is due
to a significative fraction of clusters (84 out of 309) with functional
overlap strictly equal to one.}
\label{fig:overlaps}
\end{center}
\end{figure}

We define then the cumulative distribution functions as
\begin{eqnarray}
\Pi^{(\mathrm{intra})} (O) &=& \int_0^O  P^{(\mathrm{intra})}(x)\,dx\\  
\Pi^{(\mathrm{inter})} (o) &=& \int_0^o  P^{(\mathrm{inter})}(x)\,dx\,\,\,.
\end{eqnarray}
The two cumulative functions are displayed in Fig.\ref{fig:overlaps}.
The algorithm can be run separately and in parallel on all connected
components of ${\cal U}$, because there no exchange of information
between them. Equivalently speaking, the score function can be written
as a sum over all components $c$ of separated scores: 
$E = \sum_c E_c (\{ X_i \}_{i\in c})$.

\section{Results} 

We have run our algorithm solving eqs.~(\ref{eq:prob})
and (\ref{eq:M}) at several values of $\beta > \beta_c$ and for
different choices of initial conditions of populations $\{ P_i(X_i)
\}_{i=1,...,N}$. Results are always very stable with respect to
initial conditions. Instead of maximizing the Gibbs potential directly
at zero temperature we have worked at finite $\beta$ because we were
interested also in predicting functional assignments that could be
biologically allowed although not strictly maximizing the score function
(\ref{score}).  Above $\beta_c$,
the function probabilities for each protein converge on a set of
values organized in hierarchies. The probability values are
$\beta$-dependent, but not so the hierarchical structure (see
fig~(\ref{fig:gerarchie}) for an example). All results presented in
the following are therefore taken at one given high value of $\beta$
($\beta = 2$ for the DIP PPI graph and $\beta = 10$ for the {\bf U} PPI graph.
For any protein $i$ and the connected component $c$ $i$ belongs
to, we have filtered out all background noise probability values for
functions that are not present in $c$ and still have a non zero
contributions due to the form of eq.~(\ref{eq:prob}).  We have then
collected and ranked the remaining functional probabilities, following
their emerging hierarchical structure.  A list of predicted functions
for all the unclassified proteins in the {\bf U} PPI network
using MIPS $2003$ functional categories catalog is presented in the
{\bf Supplementary Table}.  The rank division is explained in
fig.(\ref{fig:gerarchie}).  In order to probe the reliability of our
algorithm, we have followed the standard procedures of Vazquez et 
al.~\cite{ALEXEI}: starting from ${\cal G}$ and a corresponding MIPS functional
annotation, we disregarded the functions of a given fraction $d$ of
classified proteins and considered them as unclassified. We have called
$d$ ``dilution'' of ${\cal G}$. If a previously classified protein is
considered unclassified we say it has been ``whitened''. With this
procedure one obtains a new larger graph ${\cal U}_d$ of unclassified
proteins, where the algorithm can be run and its ability of finding
again the erased functions can be tested.  This testing procedure is
very similar but more stringent than the {\it Leave One-Out 
Method}~\cite{OtherFUNCTION-Methods}, because it assumes as
unclassified an extensive fraction of proteins in the graph instead of
only one each time.  We repeated the procedure for both PPI networks. 
Results for a set of performance
\begin{figure}
\includegraphics[width=3in]{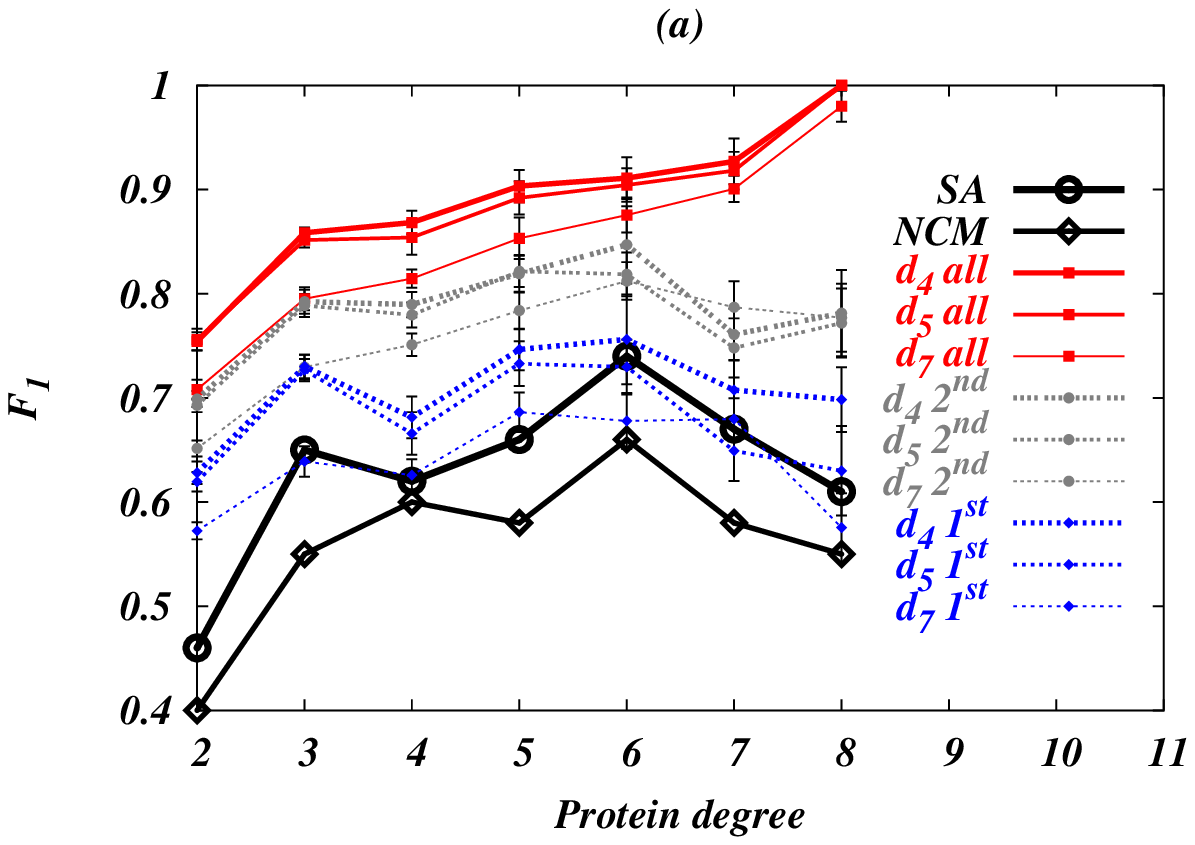}
\includegraphics[width=3in]{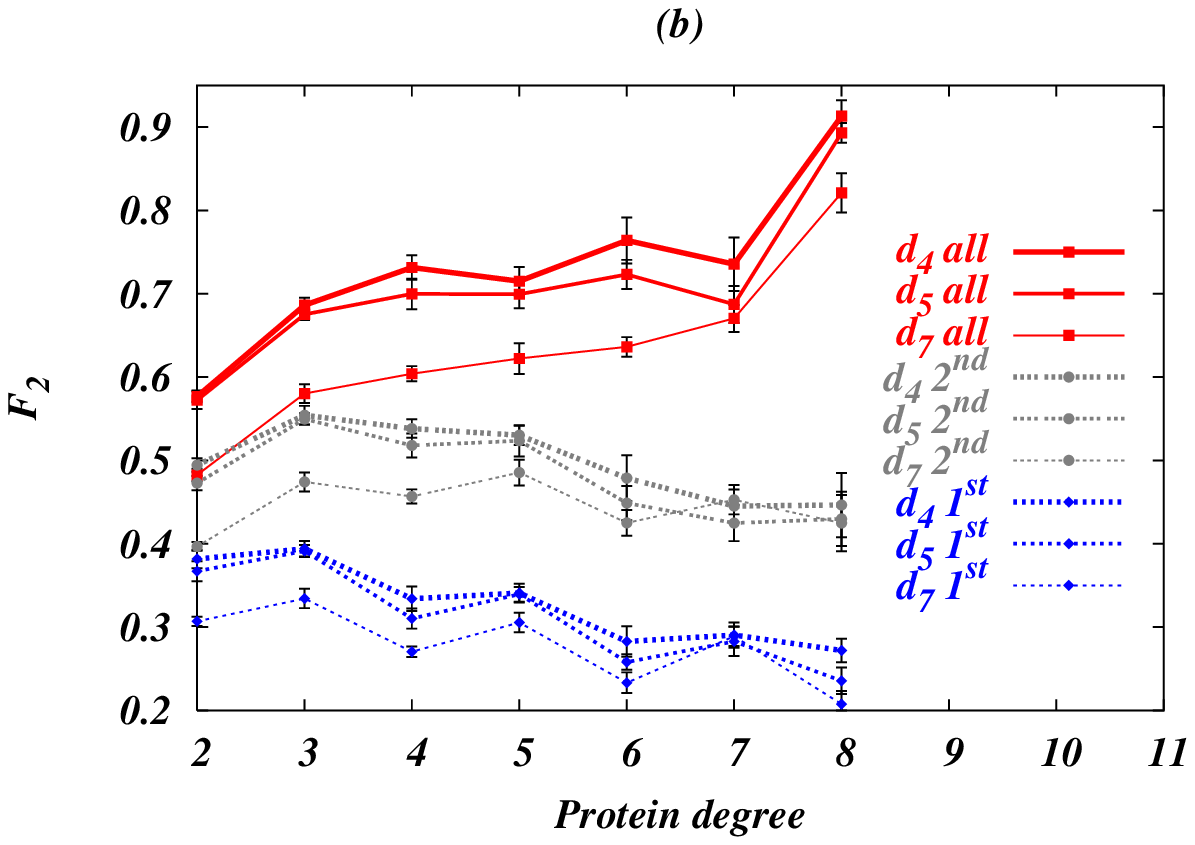}
\includegraphics[width=3in]{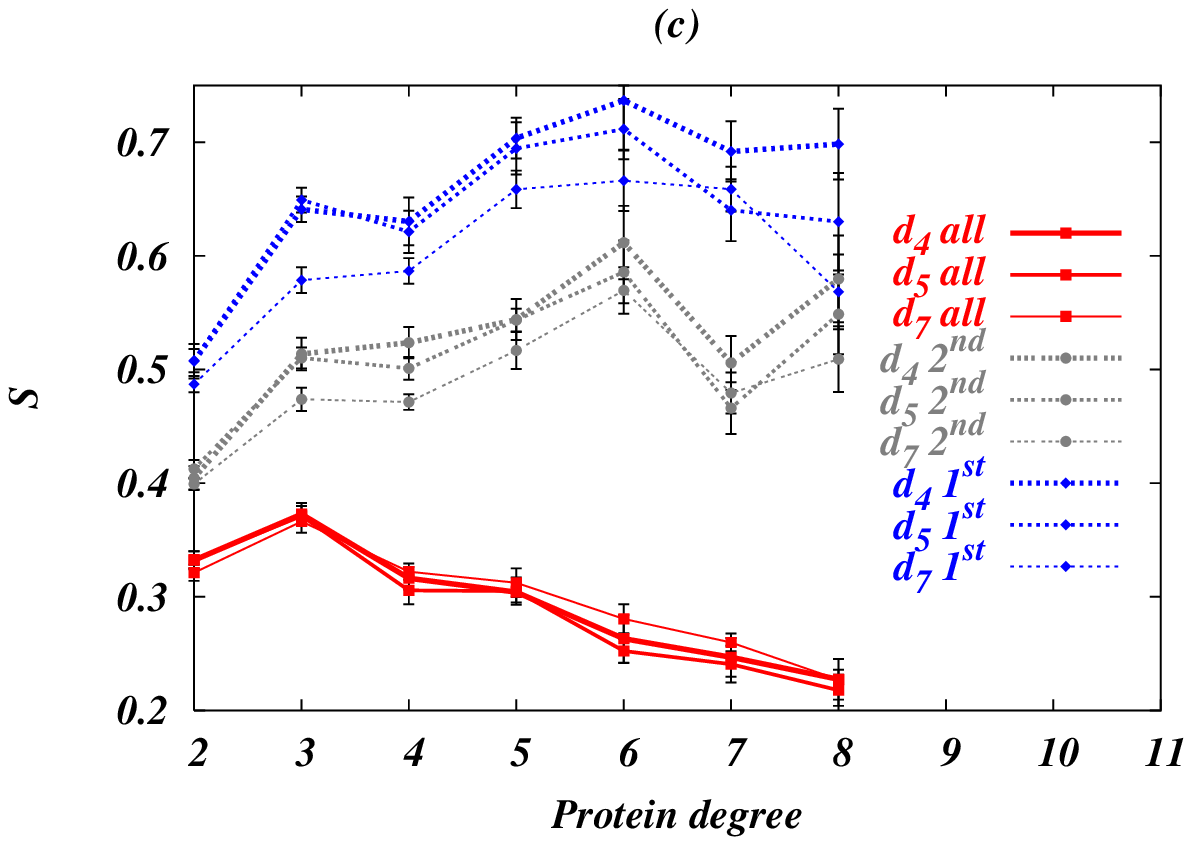}
\caption{
{\bf (a)} $F_1$, {\bf (b)} $F_2$ and {\bf
(c)} Sharpness versus protein degree for different 
{\bf U} dilutions, as described in the text.
Results are displayed for three dilution
levels $d_4 = 0.4$, $d_5 = 0.5$, $d_7 = 0.7$. Dotted lines are results
considering only functions of higher probabilities ($1^{st}$ best
rank). Dotted-dashed lines are results considering both best and second best
ranks. Thick lines consider all non background noise ranks. {\bf SA} and
{\bf NCM} are the Simulated Annealing and the
Neighboring Counting Method results for dilution $d = 0.4$. Notice that a low value of
Sharpness does not necessarily indicate a poor performance of the
algorithm. It could also be due to the fact that indeed many functions
have not been observed also in already classified proteins and
therefore the catalogs are incomplete not only for proteins in ${\cal
U}$, but on all ${\cal G}$.
}
\label{fig:tests1}
\end{figure} 
parameters are presented in fig.~(\ref{fig:tests1}) as a function of 
protein degrees for some fixed dilution values:
{\bf Fig.(a)}:{\bf Reliability}. We have defined as a first
reliability parameter $F_1$ the fraction of whitened proteins for
which the algorithm predicts correctly {\it at least one}
function. 
{\bf Fig.(b)} measures a second reliability parameter $F_2$, defined
as the fraction of correctly predicted functions out of all functions
a whitened protein has on the original graph ${\cal G}$. This test is
more stringent because it checks the ability of the algorithm of
predicting not only one function, but as many as it can. It is worth
noticing that under the $F_2$ test the method still performs very well
when all non background noise ranks are considered.  The {\it legenda}
is the same as in picture {\bf (a)}. {\bf (Fig.c)}: {\bf Sharpness}. $S$
measures the precision of the method and it is defined as the fraction
of the number of correctly predicted functions over the number of all
predicted ones. It is intuitive that the sharpness decreases with the
number probability levels (ranks) one accepts as significant. For
whitened proteins of degree 5, for instance, on average only 31$\%$ of
predicted functions belong to the set of already known erased ones,
while in the case of best ranks only, this percentage raises to
65$\%$-70$\%$. In case one allows the algorithm to predict still
experimentally unobserved functions also on the classified proteins in
${\cal G}$, the sharpness still decreases. 
\begin{figure}
\includegraphics[width=3in]{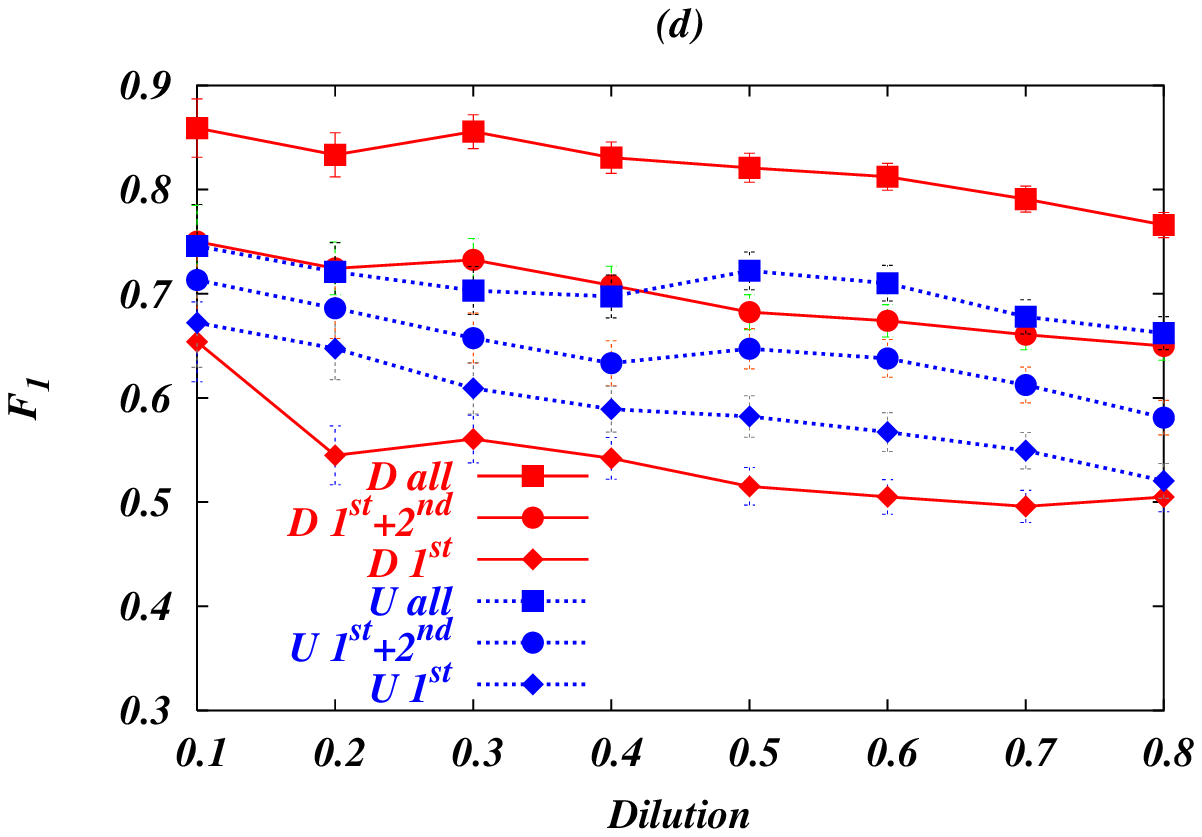}
\includegraphics[width=3in]{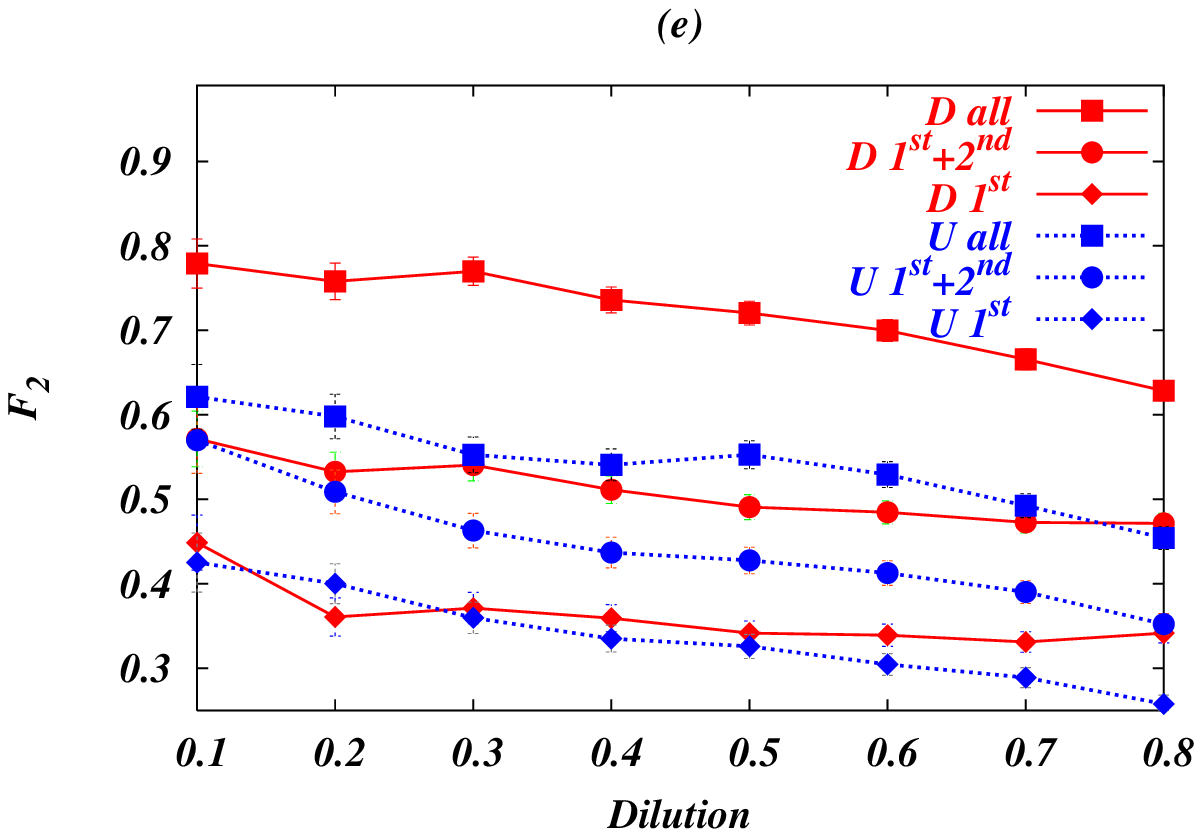}
\includegraphics[width=3in]{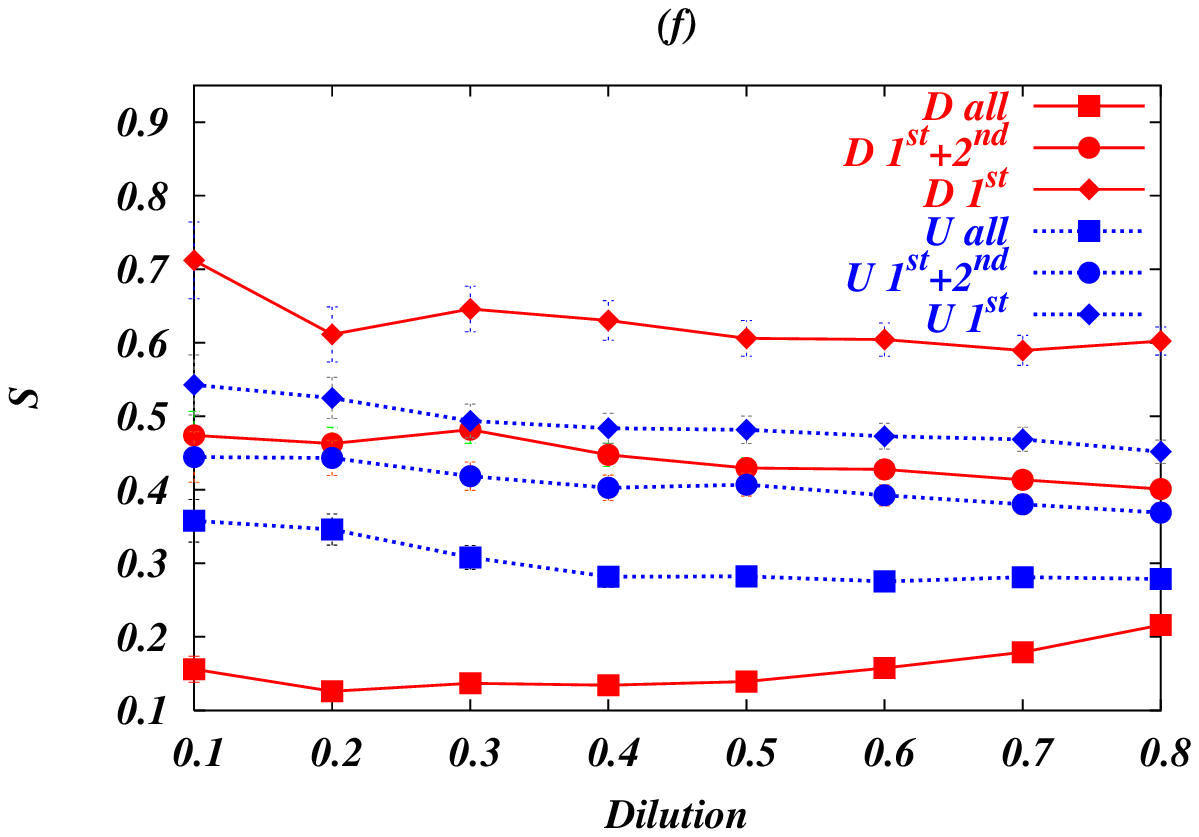}
\caption{ 
{\bf (d)} $F_1$, {\bf (e)} $F_2$ and {\bf
(f)} Sharpness versus dilution, averaged over all the PPI network and
over $n=10$ random dilution realizations. Thick lines are results for
the {\bf D} network, dotted for {\bf U}. For each network we
have again considered $1^{st}$ best, $1^{st}$ and $2^{nd}$ best and
all non noise ranks results. The different spacing between lines
comparing the two networks reflect their different topological
structure.  
Proteins to be whitened were chosen randomly in $A$. The
procedure was repeated $n = 10$ times (Larger $n$ datasets can be
easily produced, but data are already very stable for $n = 10$) for
each $d$, and the results averaged. 
We disregarded as statistically non significant
the few observed proteins with $k > 8$.
}
\label{fig:tests2}
\end{figure}
Results as a function of network dilution. 
are presented in fig.~(\ref{fig:tests2})
When (see
fig.~(\ref{fig:tests1}).(a)) a direct comparison with other methods on
the same ${\cal G}$ and MIPS catalog was possible, results of our
method were systematically better than both the Neighboring Counting
Method and the SA.  Performance further improves if we consider non
only highest rank predictions, but all significant non background
noise probabilities.  Together with the other available methods, BP
performs worse in predicting functions on leaves of the PPI graph,
i.e. on whitened proteins with only one neighbor. Nevertheless, even
in this case we observed better reliability results.

\section{Discussion}

\begin{figure}
\includegraphics[width=3in]{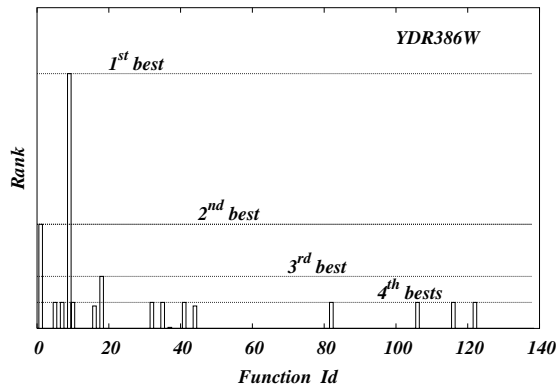} 
\caption{Example of predicted probabilities ranks for protein YDR386W
in the MIPS catalog and for the {\bf U} PPI network. In this
example, out of all possible 140 functions only the ones with a
vertical bar have non background noise probabilities. Bars heights
({\it ranks}) are proportional to the logarithm of the probability of
having a given function for all the functions ordered on the
horizontal axis.}
\label{fig:gerarchie}
\end{figure}

{\bf Hierarchical probabilities structure:} Let consider as a simple
example a protein $i$ surrounded by $3$ classified neighbors, two
having function $\sigma$ and one having function $\tau$. according to
(\ref{eq:prob}), in the zero temperature ($\beta \to \infty$) limit
one has $P_i(\sigma) = 1$ and $P_i(\tau) = 0$ together with all other
function probabilities. However, if the interaction between protein
$i$ and the one neighbor with function $\tau$ is correct, a
biologically more sound functional assignment would be that of giving
to protein $i$ both functions. Working at finite $\beta$ one can see
again from (\ref{eq:prob}) that a non zero value of $P_i(\tau)$ is
also found. The numerical value will continuously depend of
$\beta$. The hierarchy of the values of the predicted probabilities
turns out to be nevertheless very stable after crossing the critical
point $\beta_c$.  One example of probabilities at convergence at a
given $\beta$ and for a randomly chosen protein is given in
fig.~(\ref{fig:gerarchie}).

\noindent
{\bf Extension to the algorithm from the unclassified proteins graph
${\cal U}$ only to all proteins in ${\cal G}$:} Looking at four
subsequent versions MIPS databases releases ($2001$, $2002$, $2003$
and $2004$ respectively, one can see that new functions
are progressively assigned to already classified proteins too, so that
an inference procedure that allows for this possibility is in
principle more complete. However, this procedure can lead to a
spreading in the values of inferred probabilities, loosing in
Sharpness. Indeed, we tested the performance of our algorithms in both
the general and the restricted case, without noticing any significant
difference in performance.  In fact, since ${\cal G}$ is significantly
more connected than ${\cal U}$ the algorithm is significantly slower
in reaching convergence of the probability values (It still requires
one single run, being therefore faster than the SA approach).
Nevertheless, the possibility that the more general case would work
significantly better under the definition of a more refined score
function, or with a more complete and reliable PPI network, or with
the extension of the method to multi-body interactions taking into
consideration larger protein complexes cannot be ruled out.  Results
shown in the body of the text have been limited for clarity to the
restricted case where inference is measured only on proteins $\in V
\backslash A$ and for the $2002$ and $2003$ MIPS catalogs, in order to
compare them with results already present in the literature. The same
algorithm could be run on the latest $2004$ MIPS catalog release with
no effort.

\noindent 
{\bf Comparison with other available methods.}  Differently from
SA~\cite{ALEXEI}, BP algorithm allows to compute directly and in a single
run all probabilities $P_i(f)$ for a given protein $i$ to be assigned
a function $f$. This is an advantage with respect to the SA approach,
where the output of a single run is one configuration only out of a
mutually exclusive set, and in order to obtain trustful probabilities
one should average over a large number of SA runs. Moreover, provided
one can trust the numerical BP results hierarchies at convergence,
some non ground state configurations that could have a biological
sound interpretation (see Methods for details) are captured in the BP
approach in a hierarchical way, while are missed by SA unless one had
time to run a number of cooling experiments of the order of $10^6$
(Compare with the $10^2$ runs reported in Vazquez et al.~\cite{ALEXEI}).  
Differently from Kasif et al.~\cite{BIOINFO}, our version of BP naturally 
converges and does
not therefore need iterations truncation. The connection of computed
probability values with the real unknown ones can be made only at
convergence of the BP iteration equations, and it is not clear how to
interpret the probability values after only a limited number (two in
Kasif et al.~\cite{BIOINFO}) of iteration steps, when one could still be in the
middle of a transient still heavily dependent on initial
conditions. Moreover, truncating the iteration after a small number of
steps means disregarding propagation of information coming  from
distant regions of the network, which is the spirit of any message
passing algorithm like BP. The method could still in principle work if
the most distant message passing nodes of any chosen node $i \in V
\backslash A$ were a few neighboring steps away.  This turns out to be
almost the case for the considered PPI networks, due to high
clusterization and function segregation of $V \backslash A$, as
described in Methods, but it is not generally true in inference
problems. In a second Bayesian approach~\cite{OtherFUNCTION-Methods}, a large number of
external parameters (one set for each function) has to be estimated
before running the Bayesian inference algorithm.  Still, the Gibbs
potential~\cite{OtherFUNCTION-Methods} could in principle be
of a more complete form, allowing for the presence of a chemical
potential-like terms (one for each function)  proportional to the
overall number of times one function is present in the graph. However,
it is not clear what the reliability of the biological significance of
a term of this type, since influence from the classified functions of
distant proteins should already be taken into account in the structure
of the message massing procedure.   Moreover, if the property of
functional segregation was true also on the complete (still unknown)
PPI network (See Barabasi et al.~\cite{BARABASI} for some indications that this might
be the case), it is not clear why a protein should have a high
probability of being classified with a certain function only because a
large group of proteins with a very frequent function existed, even
if not interacting with the protein under consideration.  In addition,
our BP method does not require keeping track of single configurations
of functions under the iterations, but only directly of probability
weights. The algorithm converges to a stable  fixed point and does not
need the definition of a measuring time window 
period~\cite{OtherFUNCTION-Methods}.  Together with the Monte Carlo approach,
our algorithm does not need previous estimation of external parameters
defining the Gibbs potential, except for the overall tuning inverse
temperature $\beta$.

\noindent
{\bf Limitations.}  Our method has got of course many
limitations. {\bf (1)} The uncertainty over the graph
structure, due to the presence of false positive and false negative
interactions. The network topology could vary greatly and the network
instead of being divided into connected components could be made of
essentially only one giant component. The degree distribution of the
network could vary, even though some authors suggest there is evidence
for a stabilization towards a scale-free like form~\cite{BARABASI}.
Attempts of healing PPI networks errors or missing links are described
in the literature~\cite{PROTEIN-VC,OtherVC-Methods}, together with a
general description of a message passing approach to network
reconstruction~\cite{JAAKKOLA}. Our algorithm could be
generalized to partially deal in parallel with these problems,
considering two sets of dynamical variables $\{ X_i \}$ and $\{ J_{ij}
\}$ instead of $\{ X_i \}$ only. Each $J_{ij}$ could then take values
in a discrete set measuring the likelihood of the interaction between
proteins $i$ and $j$ to be present as a function of reliability of the
experimental data and of the predicted functions assigned to the
proteins under consideration. The extrapolated set $\{ J_{ij} \}$
could then be taken as a starting point to calculate new function
probabilities over the whole graph using again the BP
procedure. Extensions of the method in this direction are under study
but are not presented in this paper. {\bf (2)} Pairwise interactions
in the observed PPI graph could hide a more complex hyper-graph like
structure, with more than two proteins interacting trough protein
complexes. Our algorithm is readily generalizable to these cases, but
we have not tried to test it on actual data yet. {\bf (3)} The way the
BP algorithm predicts functions on classified proteins is intrinsically
different from the way new annotations are listed in the growing
catalogs: to the authors understanding, experiments are typically run
concentrating on one or a limited number of interesting function,
while other functions could be disregarded. The inference algorithms
predictions treat all functions in the same way, so frequency of
predicted functions could differ significantly from the experimentally
observed ones. {\bf (4)} The numerical values of the probabilities can
be proved to be correct (for a given score function) only in the case
the PPI graph is strictly a tree. This is not the case for the
experimental data, where cycles are present. However, we believe that
the order of magnitude of the $P_i(X_i)$ values should be
trustful. This approximation is one of the sources of error in the
results of fig.~(\ref{fig:tests1}).  {\bf (5)} The PPI graph is
usually built as a time and space average of all processes going all
within the cell: a given protein classified for instance with functions
$1$ and $2$ could in principle interact with two other proteins at
times in the cell cycle and/or in different places. One of the
neighboring protein could then take common function $1$, while the
other could take common function $2$, in a perfectly sound
configuration. Running the BP algorithm on the averaged graph would
however lead to the prediction of both functions on both neighbors. In
this way the algorithm would loose predictive power and sharpness,
however it would still predict the correct functions with a certain
probability whose exact numerical values should again be taken ``cum
grano salis'', as already said in point (4). {\bf (6)} Different
databases information should be merged in a proper way: for instance,
the {\bf U} and {\bf D} PPI graphs are
significantly different both in overlap and in topological structure
(this point is strictly connected with (1)).  We have decided to apply
the function prediction method to both graphs separately, but in
principle it could be used on a merging of different networks that
properly weights relative reliability of interaction links. {\bf (7)} The
Kronecker delta function defines a binary distance between functions:
one link in ${\cal U}$ contributes $1$ to the total score only if the
interacting proteins have exactly the same function; $0$ otherwise.
However, The MIPS classification scheme is organized hierarchically:
some proteins have very specific functions, while others can be
classified only in a more coarse-grained functional categories.  
The choice of a binary distance is probably appropriate
if one considers only functional categories at a given hierarchical
level but it seems unsatisfactory for the total classification, where
a more complete notion of hierarchical distance between different
functional categories would be needed. In particular one would like to
have a distance that recognizes as possibly close two neighboring
proteins of functions $\sigma$ and $\tau$ in the case $\sigma$ belongs
to a very specific functional category, while $\tau$ belongs only to a
broader one that includes the first, but with no further
specification.  In this paper we have limited the method to the binary
distance score function~(\ref{score}), considering only functions at a
chosen hierarchical level in the MIPS catalogs and disregarding all
the others. In this way some information on partial knowledge of the
functions assigned to a given protein is lost. This limitation becomes particularly 
dramatic in the case of the use of catalogs that are not organized hierarchically, but in a more 
complex way such as Gene Ontology. Extensions of this method are under study.

\section{Acknowledgements}

The authors would like to thank Paolo Provero for having given us the
starting input for this work, together with Riccardo Zecchina for fruitful 
discussions and critical reading 
of the manuscript; Alexei Vazquez, Alessandro Flammini and Vittoria Colizza 
for data and discussions.

\end{document}